\documentclass[aps,preprint]{revtex4}%
\usepackage{amsfonts}
\usepackage{amsmath}
\usepackage{amssymb}
\usepackage{graphicx}%
\providecommand{\U}[1]{\protect\rule{.1in}{.1in}}
\begin{document}
\preprint{ }
\title[ ]{On a Relativistic Quark Model Description via the Fractional Nikiforov-Uvarov Method }
\author{M. Abu-Shady}
\affiliation{{\small Department of Mathematics and computer science, Faculty of Science,
Menoufia University, Egypt}}
\author{Mohammed K. A. Kaabar}
\affiliation{{\small Institute of Mathematical Sciences, Faculty of Science, Universiti Malaya, Kuala Lumpur 50603, Malaysia
}}
\author{}
\affiliation{}

\begin{abstract}
The Dirac equation plays an essential role in the relativistic quantum
systems, which is reduced to a form similar to Schrodinger equation
when a certain potential's type is selected as the Cornell potential. By
choosing the generalized fractional derivative, the fractional Nikiforov-Uvarov
method is applied as a good efficient tool. The energy eigenvalues and corresponding wave functions are obtained in the sense of fractional forms by solving Dirac equation analytically. The special case is obtained, which is compatible with the
classical model. Solving the fractional Dirac equation will open a
new path to solve and improve results in the classical relativistic quantum systems.


\end{abstract}
\volumeyear{ }
\volumenumber{ }
\issuenumber{ }
\eid{ }
\startpage{1}
\endpage{102}
\maketitle

\section{Introduction}

Fractional-order derivative is basically a natural extension of ordinary derivatives, which has become a popular research topic in applied sciences [1-4] and engineering [5, 6]. The non-locality plays a significant role in fractional derivative models. There are various definitions of fractional derivatives such as Riemann-Liouville, Caputo, and Hadamard, where each of them has advantages and disadvantages. Therefore, there is no single fractional derivative definition that can be applied for all fractional models. 
A recently proposed generalization of fractional derivative $\left[7\right]$, named Abu-Shady--Kaabar fractional derivative (AKFD), obtains results that are compatible with the
classical fractional definitions of Riemann-Liouville and Caputo. This definition has been given in detail along with all its related theorems in $\left[7, 24\right]$, which can be written as follows:
\begin{equation}
\mathcal{D}^{AKFD}h(t)=\lim_{\Lambda\rightarrow0}\frac{h(t+\frac{\Gamma(\beta)}%
{\Gamma(\beta-\alpha+1)}\Lambda t^{1-\alpha})-h(t)}{\Lambda};\beta
>-1,\beta\in R^{+}\bigskip\tag{1}%
\end{equation}
In Eq. (1), the AKFD operator, denoted by $D^{AKFD}$, is of order $\alpha$.
The field of relativistic quantum mechanics is significant and outstanding. The
 relativistic particles' description with spin (1/2) as in [8] is heavily
dependent on the Dirac equation. The Dirac equation has been solved in recent
years for various potentials, as seen in literature [9--15]. The analytical
solutions of the Cornell potential with identical scalar and vector potentials
to the Dirac equation are examined in [9,10] using the perturbation
method and ansatz approach. For the modified-Hylleraas potential under
spin and pseudo spin symmetry limitations, the bound energy spectrum and
corresponding generalised hypergeometric wave function of the Dirac equation
are found in [5] using the Alhaidari formalism. The Nikiforov-Uvarov
approach is used in [11--13] to derive analytically the
Dirac equation's solutions with pseudoscalar Cornell, Hartmann, and
extended Hulthen potentials. Xian-Quen et al. [12] solved the Dirac equation
by using a new ring-shaped non-spherical harmonic oscillator where the scalar and vector potentials are equivalent.
The Dirac equation for a relativistic electron is generally known as one of the fundamental elements of both Standard Model (SM) and quantum electrodynamics (QED) [14,15]. Both of them are employed to look for a new physical meaning beyond the SM and to assess the viability of SM. Although the Dirac
equation was intended to explain electrons, it also successfully represents
muons, tayons, and quarks. It is astounding that the Dirac equation (DE) can
accurately describe quarks given that they are not visible in our physical
space and time. DE is utilized in phenomenological hadron models and quantum
chromodynamics (QCD) [15--20] to estimate the characteristics of heavy and
light baryons, where there are discrepancies between theoretical predictions
and experimental facts derived from quark contributions and from solving the Dirac equation

The objective of this study is to provide the fractional Dirac model by
simplifying the DE to match the Schrodinger equation. By using the
fractional form of DE, several solutions are provided for the
physical difficulties that have been described above.

The following is how the paper is arranged: The fractional Nikiforov-Uvarov (NU)
approach, which is the method employed in the current research, is briefly
presented in Section 2. Both of energy eigenvalues and wave functions of the fractional DE are computed in Section 3. The summary and conclusion are reported in Section 4.

\section{The Fractional NU (FNU) method}

This section provides a brief explanation of the FNU method for solving the
fractional differential equation in the sense of AKFD (see [7, 21] for more details), which is written as follows:
\begin{equation}
\mathcal{D}^{\alpha}\left[  \mathcal{D}^{\alpha}\Upsilon(s)\right]  +\frac{\bar{\nu}(s)}{\mu
(s)}\mathcal{D}^{\alpha}\Upsilon(s)+\frac{\tilde{\mu}(s)}{\mu^{2}(s)}\Upsilon(s)=0,
\tag{2}%
\end{equation}
where $\mu(s)$ and $\tilde{\mu}(s)$ are the polynomials of maximum $2$nd
degree of $\alpha,2\alpha,$ respectively, and $\bar{\nu}(s)$ is a polynomial
of maximum degree of $\alpha$

where%
\begin{equation}
\mathcal{D}^{\alpha}\Upsilon(s)=Is^{1-\alpha}\Upsilon^{\prime}(s), \tag{3}%
\end{equation}

\begin{equation}
\mathcal{D}^{\alpha}\left[  \mathcal{D}^{\alpha}\Upsilon(s)\right]  =I^{2}\left[  \left(
1-\alpha\right)  s^{1-2\alpha}\Upsilon^{\prime}(s)+s^{2-2\alpha}\Upsilon
^{"}(s)\right]  . \tag{4}%
\end{equation}
where%
\[
I=\frac{\Gamma\left(  \beta\right)  }{\Gamma\left(  \beta-\alpha+1\right)  }%
\]
where $0<\alpha\leq1$ and $0<\beta\leq1$. We substitute Eqs. (3) and (4) into Eq. (2) to have:%

\begin{equation}
\Upsilon^{\prime\prime}(s)+\frac{\bar{\nu}_{h}(s)}{\mu_{h}(s)}\Upsilon^{\prime
}(s)+\frac{\tilde{\mu}_{h}(s)}{\mu_{h}^{2}(s)}\Upsilon(s)=0, \tag{5}%
\end{equation}
where $\bar{\nu}_{h}(s)=\left(  1-\alpha\right)  s^{-\alpha}\mu
(s)+I^{-2}\bar{\nu}(s); \mu_{h}(s)=s^{1-\alpha}\mu(s); \tilde{\mu}%
_{h}(s)=I^{-2}\tilde{\mu}(s).$\\
To find the particular solution of Eq. (5) by
separation of variables, we deal with the following transformation:%
\begin{equation}
\Upsilon(s)=\Theta(s)\Xi(s), \tag{6}%
\end{equation}
which reduces to the following hypergeometric-type equation:%
\begin{equation}
\mu_{h}(s)\Xi^{\prime\prime}(s)+\nu_{h}(s)\Xi^{\prime}(s)+\Omega
\Xi(s)=0, \tag{7}%
\end{equation}
where%
\begin{equation}
\mu_{h}(s)=\Pi_{h}(s)\frac{\Theta(s)}{\Theta^{\prime}(s)}, \tag{8}%
\end{equation}%
\begin{equation}
\nu_{h}(s)=\bar{\nu}_{h}(s)+2\Pi_{h}(s);~\ \ \nu_{h}^{\prime}(s)<0, \tag{9}%
\end{equation}
and
\begin{equation}
\Omega=\Omega_{m}=-m\nu_{h}^{\prime}(s)-\frac{m(m-1)}{2}\mu_{h}%
^{\prime\prime}(s),m=0,1,2,... \tag{10}%
\end{equation}
$\Xi(s)=\Xi_{m}(s)$ is a polynomial of $m$ degree, satisfying Eq. (7), which is written as:%
\begin{equation}
\Xi_{m}(s)=\frac{N_{m}}{\varrho_{m}}\frac{d^{m}}{ds^{m}}(\mu_{h}%
^{\prime\prime}(s)\varrho(s)), \tag{11}%
\end{equation}
where the normalization constant is $N_{m}$, and the wight function is $\varrho(s)$ that satisfies:
\begin{equation}
\frac{d}{ds}\eta(s)=\frac{\nu(s)}{\mu_{h}(s)}\eta(s);\ \ \ \eta
(s)=\mu_{h}(s)\varrho(s), \tag{12}%
\end{equation}%
\begin{equation}
\Pi_{h}(s)=\frac{\mu_{h}^{\prime}(s)-\bar{\nu}_{h}(s)}{2}\pm\sqrt
{(\frac{\mu_{h}^{\prime}(s)-\bar{\nu}_{h}(s)}{2})^{2}-\tilde{\mu}%
_{h}(s)+\aleph\mu_{h}(s),} \tag{13}%
\end{equation}
and
\begin{equation}
\Omega=\aleph+\Pi_{h}^{\prime}(s), \tag{14}%
\end{equation}
where the polynomial of $1$st degree is $\Pi_{h}(s)$. 
The $\aleph$ value in Eq. (13) can possibly be calculated if we have square of expressions under the square root of Eq. (13). This computation can be done if we have a discriminant of zero.

\section{The DE for the Cornell Potential}

The quark mass, represented by $\omega_{\alpha}$ with index $\alpha$ which indicates the particle's kind where in our study: $u$, $d$, and $s$ quark in the presence of a confining
potential $\mathcal{V}(r)$ in the DE is given $\left[  22\right]  $ \
\begin{equation}
\left[  \mathbf{\alpha}\cdot\mathbf{\mathcal{P}+}\beta \omega_{\alpha}+\frac{1}{2}%
(1+\beta)\mathcal{V}(r)\right]  \Upsilon(\mathbf{r})=\mathcal{E}_{\alpha m}\Upsilon(\mathbf{r}), \tag{15}%
\end{equation}
where the usual DE matrices are denoted by both $\mathbf{\alpha}$ and $\beta$. Then, we decompose Eq. (15) in spherical coordinates. The following is the obtained wave function component:
\begin{equation}
\Upsilon(\mathbf{r})=\left(  _{\Theta(\mathbf{r})}^{\Xi(\mathbf{r})}\right)
=\frac{1}{r}\left(  _{iv(r)\Psi_{-q}^{n}\left(  \Phi,\Theta\right)
}^{u(r)\Psi_{q}^{n}\left(  \Phi,\Theta\right)  }\right)  \text{,} \tag{16}%
\end{equation}
We substitute Eq. (16) into Eq. (15) and then separate its radial parts. Then, we get:%
\begin{equation}
\frac{du(r)}{dr}=-\frac{q}{r}u(r)+\left[  \mathcal{E}_{\alpha m}+\omega_{\alpha}\right]
v(r), \tag{17}%
\end{equation}

\begin{equation}
\frac{dv(r)}{dr}=\frac{q}{r}v(r)-\left[  \mathcal{E}_{\alpha m}-\omega_{\alpha}-\mathcal{V}(r)\right]
u(r), \tag{18}%
\end{equation}
where the $\mathcal{E}_{\alpha m}$ is an energy eigenvalue of particle $\alpha$ and quantum number $q$, which is related to the total angular momentum
quantum number $\kappa$ as follows:
\begin{equation}
q=\left\{  _{\left(  \kappa+\frac{1}{2}\right)
=l\ \ \ \ \ \ \ \ \ \ \ \ \ \ \ \ \ \ \ \ \text{ if }~\kappa=l-\frac{1}{2}%
}^{-\left(  \kappa+\frac{1}{2}\right)  =-(l+1)\ \ \ \ \ \ \ \text{ if }%
~\kappa=l+\frac{1}{2}}\right.  \tag{19}%
\end{equation}
Eqs. (17) and (18) are reduced to a like Schrodinger equation as in $\left[
22\right]  $ as follows%

\begin{equation}
\lbrack\frac{d^{2}}{dr^{2}}-\frac{q(q+1)}{r^{2}}+2\varepsilon_{0}%
(\varepsilon_{1}-\mathcal{V}(r))]u(r)=0\tag{20}%
\end{equation}
where $\varepsilon_{0}=\frac{1}{2}\left(  \mathcal{E}_{\alpha m}+\omega_{\alpha}\right)  $
and $\varepsilon_{1}=\left(  \mathcal{E}_{\alpha m}-\omega_{\alpha}\right)  $. Let us take the
Cornell potential as follows:
\begin{equation}
\mathcal{V}(r)=fr-\frac{g}{r},\tag{21}%
\end{equation}
where arbitrary positive constants $f$ and $g$ are used. The potential has
distinctive features, including a $1$st term that represents the confinement
force and a $2$nd term that characterizes the Colombo force. Eq. (21) is
substituted into Eq. (20) to yield
\begin{equation}
\lbrack\frac{d^{2}}{dr^{2}}+2\varepsilon_{0}(\varepsilon_{1}-fr+\frac{g}%
{r}-\frac{q(q+1)}{2\varepsilon_{0}r^{2}})]u(r)=0.\tag{22}%
\end{equation}
We suppose that $r=\frac{1}{x}$, and  the characteristic radius of the meson is presented by $r_{0}$.
Therefore, this scheme depends on expanding $\frac{1}{x}$ in a power series around $r_{0},$ i. e. around $\delta=\frac{1}{r_{0}}$ in $x$ space (see [2] for more details). Eq. (22) can be written as:
\begin{equation}
\lbrack\frac{d^{2}}{dx^{2}}+\frac{2x}{x^{2}}\frac{d}{dx}+\frac{2}{x^{4}%
}(-\mathcal{D}_{1}+\mathcal{D}_{2}x-\mathcal{D}_{3}x^{2})]u(x)=0,\tag{23}%
\end{equation}
where, $\mathcal{D}_{1}=-\varepsilon_{0}(\varepsilon_{1}-\frac{3f}{\delta})$,
$\mathcal{D}_{2}=\varepsilon_{0}(\frac{3f}{\delta^{2}}+g)$, and $\mathcal{D}_{3}=\varepsilon
_{0}(\frac{f}{\delta^{3}}+\frac{q(q+1)}{2\varepsilon_{0}})$. To transfer Eq.
$\left(  23\right)  $ to the fractional form, one uses dimensionless form by
taking $y=Ax$ where $A$ equals 1 GeV.%

\begin{equation}
\lbrack\frac{d^{2}}{dy^{2}}+\frac{2y}{y^{2}}\frac{d}{dy}+\frac{2}{y^{4}%
}(-\mathcal{D}_{11}+\mathcal{D}_{22}y-\mathcal{D}_{3}y^{2})]R(y)=0, \tag{24}%
\end{equation}
where
\begin{equation}
\mathcal{D}_{11}=\frac{\mathcal{D}_{1}}{A^{2}},\mathcal{D}_{22}=\frac{\mathcal{D}_{2}}{A}. \tag{25}%
\end{equation}
From $\left[  22\right]  ,$ Eq. 24 can be written as:%

\begin{equation}
\lbrack \mathcal{D}^{\alpha}\left[  \mathcal{D}^{\alpha}R(y)\right]  +\frac{2y^{\alpha}%
}{y^{2\alpha}}\mathcal{D}^{\alpha}R(y)+\frac{2}{y^{4\alpha}}(-\mathcal{D}_{11}+\mathcal{D}_{22}y^{\alpha
}-\mathcal{D}_{3}y^{2\alpha})]R(y)=0. \tag{26}%
\end{equation}
We substitute Eqs. (3) and (4) into Eq. (26) to get:%
\begin{equation}
R^{\prime\prime}(y)+\frac{\bar{\nu}_{h}(y)}{\mu_{h}(y)}R^{\prime}%
(x)+\frac{\tilde{\mu}_{h}(y)}{\mu_{h}^{2}(y)}R(y)=0, \tag{27}%
\end{equation}
where%

\begin{equation}
\bar{\nu}_{h}(s)=\left(  1-\alpha\right)  y^{\alpha}+2I^{-2}y^{\alpha}%
,\mu_{h}(s)=y^{\alpha+1},\text{and }\tilde{\mu}_{h}(y)=2I^{-2}%
(-\mathcal{D}_{1}+\mathcal{D}_{2}y^{\alpha}-\mathcal{D}_{3}y^{2\alpha}). \tag{28}%
\end{equation}
Thus, Eq. (27) satisfies Eq. (5).\\
By substituting Eq. (28) into Eq. (13), we have: 
\begin{equation}
\Pi_{h}=-y^{\alpha}+\alpha I^{-2}y^{\alpha}\pm\sqrt{\left(  -y^{\alpha}+\alpha
I^{-2}y^{\alpha}\right)  ^{2}-2I^{-2}(-\mathcal{D}_{11}+\mathcal{D}_{22}y^{\alpha}-\mathcal{D}_{3}%
y^{2\alpha})+\aleph y^{1+\alpha}}. \tag{29}%
\end{equation}
The constant $\aleph$ is selected in a way that there is a function under the square root which has a
double zero, where its discriminant equals $0$. Therefore, we obtain:
\begin{equation}
\aleph=\left(  \frac{I^{-2}\mathcal{D}_{22}^{2}}{2\mathcal{D}_{11}}-(1-2\alpha I^{-2}+\alpha^{2}%
I^{-4}+2I^{-2}\mathcal{D}_{3})\right)  y^{\alpha-1}\text{.} \tag{30}%
\end{equation}
We substitute Eq. (30) into Eq. (29) to obtain:%

\begin{equation}
\Pi_{h}\left(  y\right)  =-y^{\alpha}+\alpha I^{-2}y^{\alpha}+\frac{\mathcal{D}_{22}%
}{\sqrt{2\mathcal{D}_{11}}}y^{\alpha}-\sqrt{2\mathcal{D}_{11}} \tag{31}%
\end{equation}
From $\left[  23\right]  $, the Eq. (31) positive sign is found. \\
From Eq. (9), we get:%
\begin{equation}
\nu_{h}(y)=\left(  1-\alpha\right)  y^{\alpha}+2y^{\alpha}-2\left(
\frac{\mathcal{D}_{22}}{\sqrt{2\mathcal{D}_{11}}}y^{\alpha}-\sqrt{2\mathcal{D}_{11}}\right)  , \tag{32}%
\end{equation}
and from Eq. (10), we have:
\begin{equation}
\Omega_{m}=\left(  -m\left(  3\alpha-\alpha^{2}\right)  -\frac{2m\mathcal{D}_{22}%
\alpha}{\sqrt{2\mathcal{D}_{11}}}-\frac{m\left(  m-1\right)  \alpha\left(
\alpha+1\right)  }{2}\right)  y^{\alpha-1}\text{.} \tag{33}%
\end{equation}
According to Eq. (10), $\Omega=\Omega_{m}$. Therefore, the Eq. (33) energy eigenvalues in the $N$-dimensional space is written as:%

\begin{align}
\mathcal{E}_{mL}^{M}  &  =\frac{3f}{\delta}-\frac{2\varepsilon_{0}I^{-4}(\frac{3f}{\delta^{2}}+g)^{2}}{[\left(
2m+1\right)  \alpha+\sqrt{\left(  2m+1\right)  ^{2}\alpha^{2}-4I^{-2}W}]^{2}}.
\tag{34}%
\end{align}
with%
\begin{equation}
W=m\left(  3\alpha-\alpha^{2}\right)  +\frac{1}{2}m\left(  m-1\right)
\alpha\left(  \alpha+1\right)  -(1-2\alpha I^{-2}+\alpha^{2}I^{-4}%
+2I^{-2}\mathcal{D}_{3})-\alpha+\alpha^{2}I^{-2}. \tag{35}%
\end{equation}
The wave function's radial is written as follows:
\begin{equation}
R_{mL}\left(  r^{\alpha}\right)  =C_{mL}~r^{\left(  -\frac{\mathcal{D}_{2}}{\sqrt
{2\mathcal{D}_{1}}}-1\right)  \alpha}e^{\sqrt{2\mathcal{D}_{1}}r^{\alpha}}(-r^{2\alpha}\mathcal{D}^{\alpha
})^{m}(r^{\left(  -2m+\frac{\mathcal{D}_{2}}{\sqrt{2\mathcal{D}_{1}}}\right)  \alpha}%
e^{-2\sqrt{2\mathcal{D}_{1}}r^{\alpha}}). \tag{36}%
\end{equation}
The normalization constant is denoted by $C_{mL}$ which is computed by $\int\left\vert
R_{mL}\left(  r^{\alpha}\right)  \right\vert ^{2}dr=1$. Eq. (36) is not explicitly dependent on the dimensions' number.
$\int\left\vert R_{mL}\left(  r\right)  \right\vert
^{2}dr=1$ stays the same without any change. Therefore, Eq. (36) is reduced to a special
case in $\left[  23\right]  $ at $\alpha=\beta=1$ and $c=0$
\ \ \ \ \ \ \ \ \ \ \ \ \ \ \ \ \ \
\begin{equation}
\varepsilon_{1}=\frac{3f}{\delta}-\frac{2\varepsilon_{0}(\frac{3f}{\delta^{2}%
}+g)^{2}}{[(2m+1)\pm\sqrt{1+\frac{8\varepsilon_{0}f}{\delta^{3}}+4q(q+1)}%
]^{2}}\text{.} \tag{37}%
\end{equation}

\section{Summary and Conclusion}

The AKFD, which has more benefits than other classical fractional derivatives, is used in connection with the fractional Nikiforov-Uvarov approach to solve the relativistic Dirac equation after being reduced to a like the Schrodinger equation which incorporates the Cornell interaction potential that is used to produce the energy eigenvalues and associated wave functions in fractional forms. The ability to describe interactions over both short and long distances is well-established by the Cornell potential. The current results at $\alpha=\beta=1$ are used as a special case to generate the classical results. The use of the Nikiforov-Uvarov approach and the AKFD in the current work has the benefit that the
fractional solution of Dirac is not taken into account in many recent
publications. In addition, the conclusion from the current study is crucial for
enhancing and resolving those from classical relativistic quantum systems. 

\section{References}

\begin{enumerate}
\item M. Abu-Shady, Sh Y. Ezz-Alarab, Conformable Fractional of the
Analytical Exact Iteration Method for Heavy Quarkonium Masses Spectra, Few-Body Systems 62, 1, (2021).

\item M. Abu-Shady, Quarkonium masses in a hot QCD medium using conformable fractional of the Nikiforov--Uvarov method, International Journal of Modern Physics A 34, 1950201 (2019).

\item M. Abu-Shady, E. M. Khokha, T. A. Abdel-Karim, The generalized fractional NU method for the diatomic molecules in the Deng--Fan model, The European Physical Journal D 76.9 (2022): 159.

\item M. Abu-shady, H. M. Fath-Allah, Masses of Single, Double, and
Triple Heavy Baryons in the Hyper-Central Quark Model by Using GF-AEIM, Advances in High Energy Physics 2022 (2022).

\item A. D. Alhaidari, H. Bahlouli, A. Al-Hasan, Phys. Lett. A
\textbf{349}, 87 (2006).

\item L. A. Trevisan, Carlos Mirez, F. M. Andrade, Few-body Syst.
\textbf{55} 1055 (2014).

\item M. Abu-Shady, M. K. A. Kaabar, A Generalized Definition of the
Fractional Derivative with Applications, Mathematical Problems in Engineering,
vol. 2021, 9444803, (2021).

\item H. Hassanabadi, E. Maghsoodi, Akpan N. Ikot, S. Zarrinkamar, Adv.
High Energ. Phys. \textbf{2014}, 831938 (2014).

\item A. N. Ikot, E. Maghsoodi, O. A. Awoga, S. Zarrinkamar, H. Hassanabadi,
Quant. Phys. Lett. \textbf{3}, 7 (2014).

\item M. Hamzavi, A. A. Rajabi, Chinese Phys. C \textbf{37}, 103102 (2013).

\item M. Hamzavi, H. Hassanabadi, A. A. Rajabi, Inter. J. Mod. Phys. E,
19, 2189 (2010).

\item H. Xian-Quen, L. Guoug, W. Yhi-Min, N. Lion-Bin, M. Yan. Commun.
Theor. Phys. \textbf{53}, 242 (2010).

\item A. Arda, R. Sever, C. Tezcan, Cent. Eur. J. Phys. \textbf{8}, 843 (2010).

\item N.N. Bogoliubov, A.A. Logunov, A.I. Oksak, I.T. Todorov, General principles of quantum field theory. Moscow, Nauka, 1987 (in Russian).

\item F. J. Yndurain, Quantum Chromodynamics. Springer-Verlag, New York, Berlin- Heidelberg-Tokyo, 1983.

\item M. Abu-Shady, Chiral Logarithmic Quark Model of N and $\Delta$ with an A-Term in the Mean-Field Approximation, International Journal of Modern Physics A 26.02 (2011): 235-249.

\item M. Abu-Shady, Meson properties at finite temperature in the linear sigma model, International Journal of Theoretical Physics 49 (2010): 2425-2436.

\item M. Abu-Shady, M. Soleiman, The extended quark sigma model at
finite temperature and baryonic chemical potential, Physics of Particles and Nuclei Letters 10 (2013): 683-692.

\item M. Abu-Shady, Effect of logarithmic mesonic potential on nucleon properties, Modern Physics Letters A 24.20 (2009): 1617-1629.

\item M. Abu-Shady, Nucleon Properties Below the Critical Point Temperature, International Journal of Theoretical Physics 50 (2011): 1372-1381

\item H. Karayer, D. Demirhan, F. B\"{u}y\"{u}kk, Commun. Theor. Phys. 66,
12 (2018).

\item L. A. Trevisan, Carlos Mirez, F. M. Andrade, Few-body Syst.
\textbf{55} 1055 (2014).

\item M. Abu-Shady, Analytic solution of Dirac Equation for extended Cornell Potential using the Nikiforov-Uvarov method, Boson Journal of Modern Physics,
1, 1, 61, arXiv:1507.03706 (2015).

\item F. Martínez, M. K. A. Kaabar, A Novel theoretical investigation of the Abu-Shady--Kaabar fractional derivative as a modeling tool for science and engineering, Computational and Mathematical Methods in Medicine, vol. 2022, 4119082, (2022).

\end{enumerate}

\end{document}